# Composite Material Design for Optimized Fracture Toughness Using Machine Learning

Mohammad Naqizadeh Jahromi[1,*], Mohammad Ravandi [2]

[1]*Department of Mechanical and Aerospace Engineering, University of Central Florida, Orlando, USA*

[2]*Aerostructures Innovation Research Hub (AIR Hub), Swinburne University of Technology, Melbourne 3122, VIC, Australia*

*Corresponding author: mohammad.naqizadehjahromi@ucf.edu

# Composite Material Design for Optimized Fracture Toughness Using Machine Learning

**Abstract**:

This paper investigates the optimization of 2D and 3D composite structures using machine learning (ML) techniques, focusing on fracture toughness and crack propagation in the Double Cantilever Beam (DCB) test. By exploring the intricate relationship between microstructural arrangements and macroscopic properties of composites, the study demonstrates the potential of ML as a powerful tool to expedite the design optimization process, offering notable advantages over traditional finite element analysis. The research encompasses four distinct cases, examining crack propagation and fracture toughness in both 2D and 3D composite models. Through the application of ML algorithms, the study showcases the capability for rapid and accurate exploration of vast design spaces in composite materials. The findings highlight the efficiency of ML in predicting mechanical behaviors with limited training data, paving the way for broader applications in composite design and optimization. This work contributes to advancing the understanding of ML's role in enhancing the efficiency of composite material design processes.

# 1. Introduction

In modern mechanical engineering, the quest for favorable material qualities with flexible functionalities is paramount. The microstructural arrangement of materials has proven to be a significant consideration, playing a substantial role in defining the macro features of composites [1–3]. Composites, typically composed of two or more essentially distinct materials, exhibit thoroughly different large-scale features when contributing materials with various configurations are substituted [4]. The laminates of composites, formed by combining different fibers and matrices, significantly contribute to their behaviors under various loading conditions [5,6]. Traditional methods of composite manufacturing, constrained by the complexity of the gluing stage, are time-consuming. This limitation arises from the necessity to mount distinct plies on each other, with resin placed among prefabricated plies [7]. However, additive manufacturing, particularly 3D printing, has emerged as a promising tool, enabling the synthesis of composites with varying materials and features in 3D space, overcoming the hindrance of producing composites with diverse design complexities and numerous possible combinations [8].

Due to their high throughput and diverse mechanical behaviors, composites find wide application in various industrial areas. Achieving an optimized model opens the door for enhanced applicability across industries. Gu et al. dedicated efforts to investigate 2D checkerboard composite design, identifying tougher and stronger configurations [9]. In another study, they explored biomimicry in a hierarchical 2D composite, aiming to eliminate inferior configurations in terms of toughness and strength. The high-performing microstructures resulting from their research were evaluated via additive manufacturing, showcasing potential configurations among the spatial possibilities [10]. Liu conducted a survey analyzing delamination growth in a laminated composite under compression, accompanied by buckling, using the finite element method (FEM). The study also calculated the energy release rate through the virtual crack closure technique (VCCT), incorporating failure criteria such as B-K, Reeder, and power law [11]. Liu noted that notched laminated composites undergo catastrophic failure and complicated damage, including interlaminar delamination and intralaminar damage under tensile loading [12]. Numerous experimental and numerical research endeavors have explored the mechanical behaviors of 2D and 3D composites deployed across multiple industries, from remedial laminates in dentistry to aircraft wing manufacturing in the aviation industry. The structural and behavioral optimizations of composites hold significant potential for further studies.

Subsequently, it can be said that certain features of composites take precedence based on applicability. Among various features, toughness has drawn considerable attention due to its significant role in composite integrity [9,10,13]. Tuning composite toughness, such as interlaminar fracture toughness, is deemed challenging in materials' behavioral optimization [14]. Interlaminar fracture toughness exerts profound effects on a composite, and its variation leads to significant fluctuations in desired properties. For example, Jung et al. studied the behavior of glass fiber-reinforced polypropylene composites under a low-velocity impact test, considering their interlaminar fracture toughness using FEM [15].

Various methods have been proposed to investigate materials failure, especially composites failure. Since materials, especially brittle ones, often fail in the presence of cracks, this area provides a broad background for further studies [9]. Two widely used methods for predicting crack behaviors are the extended finite element method (XFEM) and VCCT. However, for modeling interlaminar fracture toughness, another solution is to employ cohesive elements, a method widely used by researchers. Yang et al. proposed a modified cohesive zone model (CZM) for the soft adhesive layer, considering the rate dependence of intrinsic fracture energy along with dissipative energy [16]. Their model's accuracy was evaluated through FEM using the double cantilever beam (DCB) debonding test, assessing the impact of intrinsic fracture energy rate dependence on cohesive zone behavior, specifically its failure. In another investigation, Zani et al. worked on determining the comprehensive constitutive response of the cohesive interface in the mode I delamination of Fully-Uncoupled Multi-Directional laminates in the DCB test, employing an energy-based approach [17].

Despite numerous works clarifying the cohesive zone mechanism and its modeling, more investigation is needed on how the cohesive zone is affected by the structural design of cantilever beams in a DCB test. Consequently, in this paper, we endeavored to study the contributing factors of the structural arrangement of composites leading to distinct behaviors performed by the cohesive zone and discern the optimized structure for our presented model. In other words, we investigated how each unique arrangement of materials plays a substantial role in the overall properties of our considered composite. However, conventional methods of optimal model diagnosis are considered significantly time-consuming due to the broad-ranging possible designs and combinations of materials.

Although finite element analysis provides efficient problem-solving in mechanical design, it is considerably handicapped in dealing with large-scale issues in terms of speed. Thus, many researchers have turned to inverse engineering approaches as an alternative to conventional fabrication methods. Machine learning (ML), a subcategory of artificial intelligence (AI), has emerged as a promising tool, enabling computers to learn from a limited set of data and predict outputs without explicit programming [18–22]. ML is considered a well-justified and feasible data-driven substitute for physics-based approaches as long as it accelerates solution modeling with acceptable accuracy. ML algorithms operate within various scientific fields, from medicine to engineering, encompassing both classification and regression applicability [23]. In 2018, Gu et al. employed an ML algorithm to generalize a checkerboard composite behavior, separating high-performing designs from low-performing ones based on their toughness and strength. They also sought to discern the sufficient amount of data for training the ML algorithm [9]. Kollmann et al. investigated a deep learning (DL) algorithm based on the convolutional neural network (CNN) to achieve optimal features, including the bulk modulus, shear modulus, and Poisson's ratio, for their metamaterial structure [24]. Li et al. have demonstrated the potential of these techniques in predicting the transverse mechanical properties of unidirectional Carbon Fiber Reinforced Polymer (CFRP) composites with microvoids [25]. Furthermore, Liu et al. demonstrated the application of ML and feature representation in accurately predicting the stress-strain curves of additively manufactured metamaterials [26]. Therefore, in this paper, we employed several ML algorithms to discern the optimized patterns of 2D and 3D composite beams in DCB tests concerning their toughness and crack propagation. Subsequently, we drew an analogy between utilized algorithms to recognize the robust approach with the highest accuracy.

## 2. Methodology

### 2.1 Computational framework

This study aims to predict and optimize the design of 2D checkerboard and 3D laminate composite cantilever beams for the DCB test, with enhanced fracture toughness as the primary objective. ML techniques are employed to analyze a limited set of possible configurations. Defining input features is crucial for utilizing ML algorithms effectively;

hence, the structural design arrangement of elements is designated as the input feature for this investigation, as described in the following sections.

A reliable sampling method is essential for training well-performing ML algorithms. To this end, a brute-force algorithm is implemented to generate all possible configurations in a predetermined sequence. Cluster sampling is then performed to create a unified subset of configurations for further finite element method (FEM) evaluation. The selected sample of configurations is analyzed using the Abaqus, and the resulting fracture toughness and crack propagation data are extracted.

Subsequently, the input features and output targets are subjected to various ML algorithms, and an analogy is drawn between the algorithms' accuracy and performance in generalizing the FEM results. Based on the best-performing algorithm, the optimal configurations for the considered target (fracture toughness) can be selected.

## 2.2 Finite element model

Two separate models, 2D and 3D, have been proposed to evaluate mechanical properties, specifically crack length propagation for the 2-D model (or crack area propagation for the 3-D model) and fracture toughness. Among these properties, fracture toughness has drawn significant attention in achieving an optimized design. The Modified Beam Theory (MBT), outlined in ASTM D5528, can be enlisted for modeling mode I interlaminar fracture toughness as below [27]:

$$G_1 = \frac{3P\delta}{2b\,(a + |\Delta|)} \qquad (1)$$

where $P$ is the opening load, $\delta$ is the crosshead displacement, $b$ is specimen width, $a$ is the crack length, and $\Delta$ is the intercept of the plot of the cube root of the specimen compliance, $\delta/P$, versus the crack length, $a$. The indicated parameters are depicted in **Figure 1**.

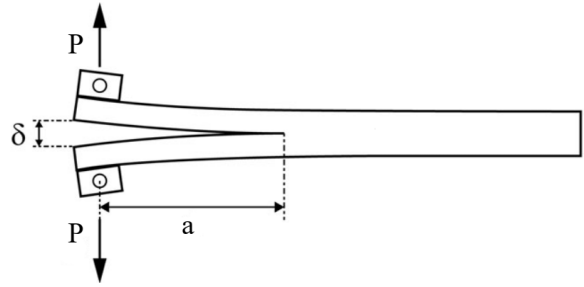

**Figure 1:** Schematic of double cantilever beams test.

Under the ASTM standard, the MBT method is generally suggested because it applies the most conservative approach to predicting fracture toughness. Hence, in this paper, having evaluated the model through MBT, FEM results were considered the fundamental truth of our investigation; more specifically, they were utilized as the input data for our ML models.

In this work, delamination is modeled by cohesive elements COH2D4 and COH3D8 for 2D and 3D models, respectively. The traction-separation model and quads damage initiation have been used to define criteria and the constitutive response of cohesive elements. BK law has been employed to describe mixed mode behavior, while energy has been set for damage evolution type. The presence of reaction forces data where the loadings were applied to and crack propagation properties is critical to calculating the fracture toughness.

*2.2.1  2-D model*

Two double cantilever beams are joined together, with a layer of cohesive elements set between them. Each beam is divided into 32 elements (arranged in a 2 by 16 grid in the y and x directions, respectively) with a 12.5% volume fraction of soft elements. In this 2-D model, four-node elements have been utilized for FEM, with degrees of freedom in the x and y directions. To streamline the configurations for further numerical investigation, a symmetric configuration for beams against each other has been defined.

The brute force algorithm suggests that 35,960 various configurations (i.e., 32C4) for our proposed 2-D composite exist.

As illustrated in **Figure 2**, an edge crack of 25% of the beam length has emerged as we joined beams and the cohesive layer to initiate crack propagation. A displacement load of 50% of the beam's width has been applied to each beam in the y-direction while the entire structure is fixed on the right side. Stiff elements (colored in red) have ten times greater elasticity modulus than soft elements (colored in green), with soft elements considered as auxetic material to alleviate crack propagation [28,29].

*2.2.2   3-D model*

As shown in **Figure 3**, the 3-D model comprises two double cantilever beams with a cohesive layer between them, similar to the 2-D model. Here, beams are assumed to be laminate, constituting five laminas in the x-y plane with a unique fiber orientation. The fibers' orientation angle is selected from our predefined set of -60, -45, -30, -15, 0, 15, 30, 45, 60, and 90 degrees from the x-axis.

Among various arrangements of fibers' orientation in the laminate, selecting five candidates among ten with respect to their order leads to 30,240 (i.e., 10P5) unique laminate designs. The arrangement in each beam is mirrored compared to the other beam to reduce calculation size, while a one-element width structure - possessing robust applicability - has been utilized to alleviate calculation as well [30]. The beams are fixed on one side, while the other is subjected to a displacement load of 150% of the beam's thickness in the z-direction. An edge crack with 30% of the beam's length is located adjacent to the cohesive layer and between the beams.

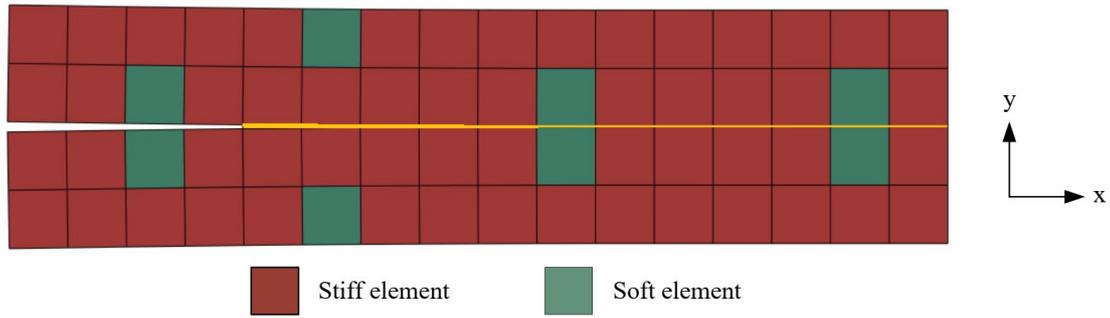

**Figure 2:** An arbitrary geometric pattern of stiff elements, soft elements, and the cohesive layer (depicted in yellow) in the 2-D model, captured shortly after applying load displacement.

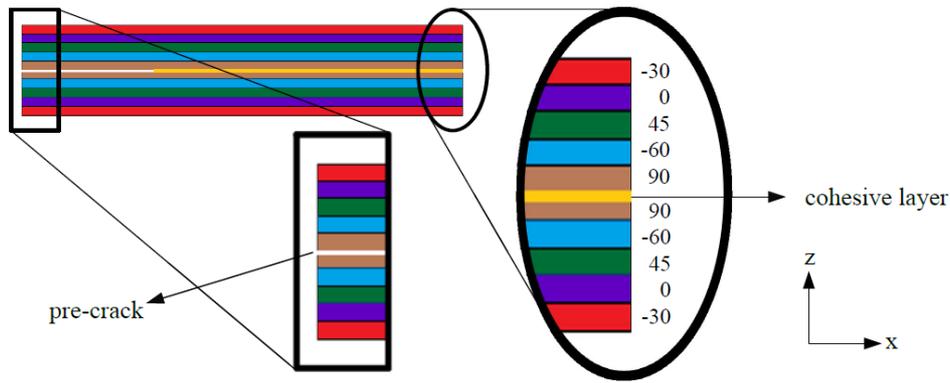

**Figure 3:** An arbitrary arrangement of laminas in the 3-D model with varying fiber orientations.

## 2.3 Machine learning approach

The scikit-learn library is employed as our general framework for deploying ML algorithms [31]. The configurations of the 2-D model (4 by 16) and 3-D model (10 by 1) composites are treated as ML input features, while the mechanical behaviors serve as target outputs. Due to symmetrical loading conditions and the design of both cases, calculations can be reduced to only half of the whole structure. Thus, 2 by 16 and 5 by 1 input features may be sufficient for the ML framework of 2-D and 3-D models, respectively. However, utilizing 5 features for the 3-D model may lead to underfitting and poor generalization during training; thus, a deeper insight and feature engineering are required. To address this, general non-zero arrays of the compliance matrix for each lamina have been used as input features. Considering the fiber orientation, the laminas can be assumed as monoclinic material, resulting in a compliance matrix with 21 non-zero and unique values (a symmetric compliance matrix holds 21 unique values).

For the 2-D model, each combination of all possibilities forms a 2 by 16 matrix, which can be flattened into an array of 32 features (i.e., 1 for the soft element and 10 for the stiff element). On the other hand, the 3-D model is considered an array of 65 features (i.e., the fiber orientation of the 5 laminas leads to 65 input features). In this work, mechanical behaviors generated by FEM are considered our target outputs in the ML framework. The prediction accuracies of various algorithms are assessed based on how well they can discern the target values of unseen geometries. Mini-batch gradient descent is employed to achieve an acceptable trade-off between training time and accuracy. In other words, the algorithms are trained with varying amounts of data, ranging from 100 to 8,000 configurations, to assess the influence of data density on prediction accuracy. Simultaneously, the test data remains unchanged (2,000 configurations) during this experiment.

The criterion used for measuring the discrepancy between the predicted value by ML and the actual value by FEM is the normalized root mean squared error (NRMSE). The equation for NRMSE is as below:

$$NRMSE = \frac{RMSE}{y_{max} - y_{min}} \quad (2)$$

where $y_{max}$ and $y_{min}$ stand for maximum and minimum values obtained by FEM, respectively. The RMSE is:

$$RMSE = \sqrt{\frac{1}{m}\sum_{i=1}^{m}(y_i - \hat{y}_i)^2} \quad (3)$$

where m is the number of configurations evaluated by ML, and $y_i$ and $\hat{y}_i$ are actual and predicted values of the $i^{(th)}$ configuration, respectively.

Another criterion widely used for evaluating ML algorithm predictions is the $R^2$ score (coefficient of determination), which indicates the proportion of variance in the target value. Therefore, it measures how well unseen samples are likely to be predicted. The equation of $R^2$ is as below:

$$R^2(y, \hat{y}) = 1 - \frac{\sum_{i=1}^{m}(y_i - \hat{y}_i)^2}{\sum_{i=1}^{m}(y_i - \bar{y}_i)^2} \tag{4}$$

where $m$ is the number of configurations evaluated by ML. $y_i$ and $\hat{y}_i$ are actual and predicted values of the $i^{(th)}$ configuration, respectively. The equation for $\bar{y}_i$ is as below:

$$\bar{y}_i = \frac{1}{m}\sum_{i=1}^{m} y_i \tag{5}$$

### 2.3.1  Linear Regression (LR)

Linear regression fits a linear model (hypothesis) with coefficients $W = (w_1, w_2, ..., w_n)$ to minimize the objective function of the equation (7) based on the targets predicted as below [31]:

For $j^{th}$ training set:

$$h_w(x^{(i)}) = w_j\, x_j^{(i)} = w_0 x_0^{(i)} + w_1 x_1^{(i)} + w_2 x_2^{(i)} + \cdots + w_n x_n^{(i)} \tag{6}$$

$$\min\ J(w_0, w_1, ..., w_n) = \frac{1}{2m}\sum_{i=1}^{m}(h_w(x^{(i)}) - y^{(i)})^2 \tag{7}$$

where $h$ is the hypothesis, $w_j$ is the $j^{(th)}$ array of the weight vector, $x_j^{(i)}$ is the value of feature $j$ in $i^{(th)}$ training example, $n$ is the number of features, $J$ is the cost function, and $m$ is the number of training sets.

*2.3.2 LASSO*

The LASSO is a linear model that estimates sparse coefficient regularization (equal to 1 and multiplied by $\alpha$). Lasso is a subcategory of the Elastic-Net algorithm, which will be elaborated on further. Here is the optimization objective for LASSO [31]:

$$\min \quad J(w_0, w_1, \ldots, w_n) = \frac{1}{2m}\sum_{i=1}^{m}(h_w(x^{(i)}) - y^{(i)})^2 + l_1 \alpha \sqrt{w_0 + w_1 + \cdots + w_n} \tag{8}$$

where $\alpha$ is the regularization term, and $l_1$ in Lasso is equivalent to 1; hence, the equation would result in:

$$\min \quad J(w_0, w_1, \ldots, w_n) = \frac{1}{2m}\sum_{i=1}^{m}(h_w(x^{(i)}) - y^{(i)})^2 + \alpha \sqrt{w_0 + w_1 + \cdots + w_n} \tag{9}$$

*2.3.3 Elastic Net (EN)*

Elastic Net is a linear regression model trained with $l_1$ and $l_2$-norm regularization of the coefficients. The elastic-Net objective function is as below [31]:

$$\min \quad J(w_0, w_1, \ldots, w_n) = \frac{1}{2m}\sum_{i=1}^{m}(h_w(x^{(i)}) - y^{(i)})^2 + l_1 \alpha \sqrt{w_0 + w_1 + \cdots + w_n} + \frac{l_2 \alpha}{2}(w_0 + w_1 + \cdots + w_n) \tag{10}$$

where $l_1$ and $l_2$ are penalty factors that can be controlled and $\alpha$ is the regularization term.

*2.3.4 K-Nearest Neighborhood Regressor (KNN)*

K-Nearest neighborhood is among the most widely used algorithms for regression and classification problems. This algorithm uses feature similarity intending to predict values for unseen data; this implies that new data are allocated a value based on how closely they resemble a data set in the training set. K-neighborhood regressor implements learning based on *k* nearest neighbors of each query point. This integer number of points must be defined manually. When using uniform weights for each point, nearby points contribute more to regression than distant points [31]. Generally, the Euclidean distance is used for the assessment of points assignment, as shown below:

$$(p,q) = \sqrt{\sum_{i=1}^{n}(q_i - p_i)^2} \qquad (11)$$

where *p* and *q* are two points in Euclidean n-space, $q_i$ and $p_i$ are Euclidean vectors starting from the origin, and *n* is Euclidean n-space.

### 2.3.5 Decision Tree Regressor (CART)

Decision Trees are non-parametric ML methods that can be implemented for classification and regression problems. This method predicts target values by interpreting rudimentary decision rules deduced from features. More specifically, it employs a set of if-then-else rules for target estimation. Among its advantages, simplicity in interpretation, visibility, and logarithmic cost make it a robust method in ML; however, slight variations in data might result in a thoroughly different generated tree. As illustrated in **Figure 4**, throughout training, decision trees recursively section features and corresponding target values in the nodes, subgrouping data with similar target values. For its mathematical interpretation, assuming $Q_m$ as the data at the node *m*, with $N_m$ being the samples; for each case split $\theta = (j, t_m)$ comprising feature *j* and the threshold $t_m$, the data is separated into distinct subsets of $Q_m^{left}(\theta)$ and $Q_m^{right}(\theta)$ as below [31]:

$$Q_m^{left}(\theta) = \{(x,y)|\ x_j \leq t_m\} \qquad (12)$$

$$Q_m^{right}(\theta) = Q_m \backslash Q_m^{left}(\theta) \qquad (13)$$

Afterward, the quality of a candidate split of the node *m* is computed by loss function

$H()$. Eventually, the parameters minimizing the impurity are selected:

$$G(Q_m, \theta) = \frac{N_m^{left}}{N_m} H\left(Q_m^{left}(\theta)\right) + \frac{N_m^{right}}{N_m} H\left(Q_m^{right}(\theta)\right) \quad (14)$$

$$\theta^* = argmin_\theta\, G(Q_m, \theta) \quad (15)$$

Recurse for both subsets, $Q_m^{left}(\theta^*)$ and $Q_m^{right}(\theta^*)$, until permissible depth is achieved, either $N_m < min_{samples}$ or $N_m = 1$. Here, we utilized CART decision tree algorithm, which constructs binary trees employing feature and the threshold that yield the most significant information gain at each unique node.

### 2.3.6 Random Forest Regressor (FOREST)

A random forest regressor is a supervised algorithm that employs an ensemble method to combines predictions from multiple algorithms (Here are decision trees) [31]. As envisioned in **Figure 4**, the random forest regressor fits several trees on various subsamples of the dataset and averages the eventual outputs to improve accuracy and mitigate over-fitting probability.

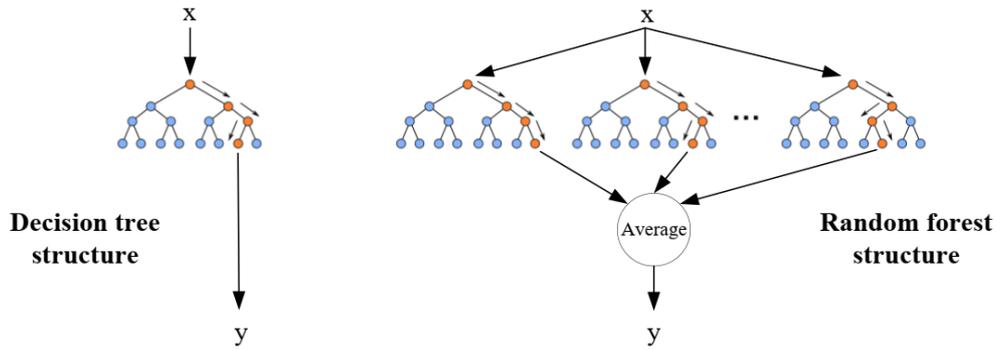

**Figure 4:** Analogy between decision tree regressor and random forest regressor structures.

### 2.3.7 Support Vector Regressor (SVR)

Support Vector Regressor (SVR) is a less commonly used but robust method that offers reliable tools for regression problems, compared to Support Vector Classification (SVC), which is designed for solving classification cases. Despite their distinct purposes, the fundamentals of both methods are identical. SVC relies on a subgroup of all training sets, located within the user-defined margin, and neglects out-of-margin data when calculating the cost function. This principle also applies to SVR, as it ignores samples with predictions close to their actual target values [31]. The SVR method constructs a hyperplane in a high or infinite-dimensional space that can be used for regression. The mathematical formulation, based on the *libsvm* library, suggests the optimization goal as below [32]:

$$\begin{aligned} min_{\omega,b,\zeta,\zeta^*} \quad & \frac{1}{2}\omega^T\omega + C\sum_{i=1}^{n}(\zeta_i + \zeta_i^*), \\ \text{Subject to} \quad & y_i - \omega^T\phi(x_i) - b \leq \varepsilon + \zeta_i, \\ & b + \omega^T\phi(x_i) - y_i \leq \varepsilon + \zeta_i^*, \end{aligned} \quad (16)$$

where training vector is defined as $x_i \in \mathbb{R}^p$, i = 1, …, n, $y \in \mathbb{R}^n$, and $\omega \in \mathbb{R}^p$, $C$ as the regularization term (penalty term), $\varepsilon$ for specifying $\varepsilon$-tube length, $\zeta$ and $\zeta_i^*$ as the distance of the point from upper or lower hyperplanes.

*2.3.8 Neural Networks (NN)*

Neural Network (NN), a non-linear and computationally intensive ML algorithm, has been employed to explore its applicability in modeling our problem using the *Keras* library [33]. Given the design complexity of our model, with 32 input features for the 2-D model and 65 input features for the 3-D model, it is essential to consider that an increase in the number of hidden layers can impact speed due to extensive calculations. Therefore, a trade-off between accuracy and speed is crucial when constructing a network structure.

In the following sections, it will be revealed that four projects must be addressed since we are dealing with two and three-dimensional composites, with crack propagation and fracture toughness as our target values. In each project, a NN structure with one hidden layer and the number of neurons equal to the number of input features is utilized, except for the three-dimensional composite with fracture toughness as the target. In this exceptional case, a deeper structure with 3 hidden layers – 150, 100, and 50 neurons (diamond structure) – and dropout layers is employed, considering the high nonlinearity of the system. The ReLU activation function and a slight bias term are applied with 100 epochs (except 400 epochs for the last problem) for training the NN model.

## 3. Results

To illustrate how the results are affected by various parameters, the impact of batch size and epochs on neural networks, as well as the impact of training data density on all algorithms, are evaluated. For this purpose, variation effect of one parameter on accuracy has been measured while keeping other parameters fixed.

### 3.1 Crack propagation of 2-D model

In this section, the ML approach targets the crack propagation of the 2-D composite. As represented in **Figure 5**, $NRMSE$ and $R^2$ for all algorithms almost stabilize after being trained by 20% of data, with a negligible gap existing between the accuracies obtained on the train and test data. Hence, these approaches successfully overcome the overfitting issue. The linear approaches – LR, LASSO, and EN – plateau between 0.92 and 0.94 for $R^2$ and around 0.045 for $NRMSE$. KNN exhibits the weakest performance among all methods and is unable to exceed 0.80 for the $R^2$. On the other hand, the remaining algorithms prove to be efficient, reaching approximately the peak of 0.98 to 0.99 for $R^2$. It is worth mentioning that CART always performs flawlessly on the training data, holding 0 and 1 values for $NRMSE$ and $R^2$, respectively. Moreover, the graph implies that the results on both training and test data for SVR and NN are nearly identical, suggesting that these methods are not susceptible to overfitting in the case study.

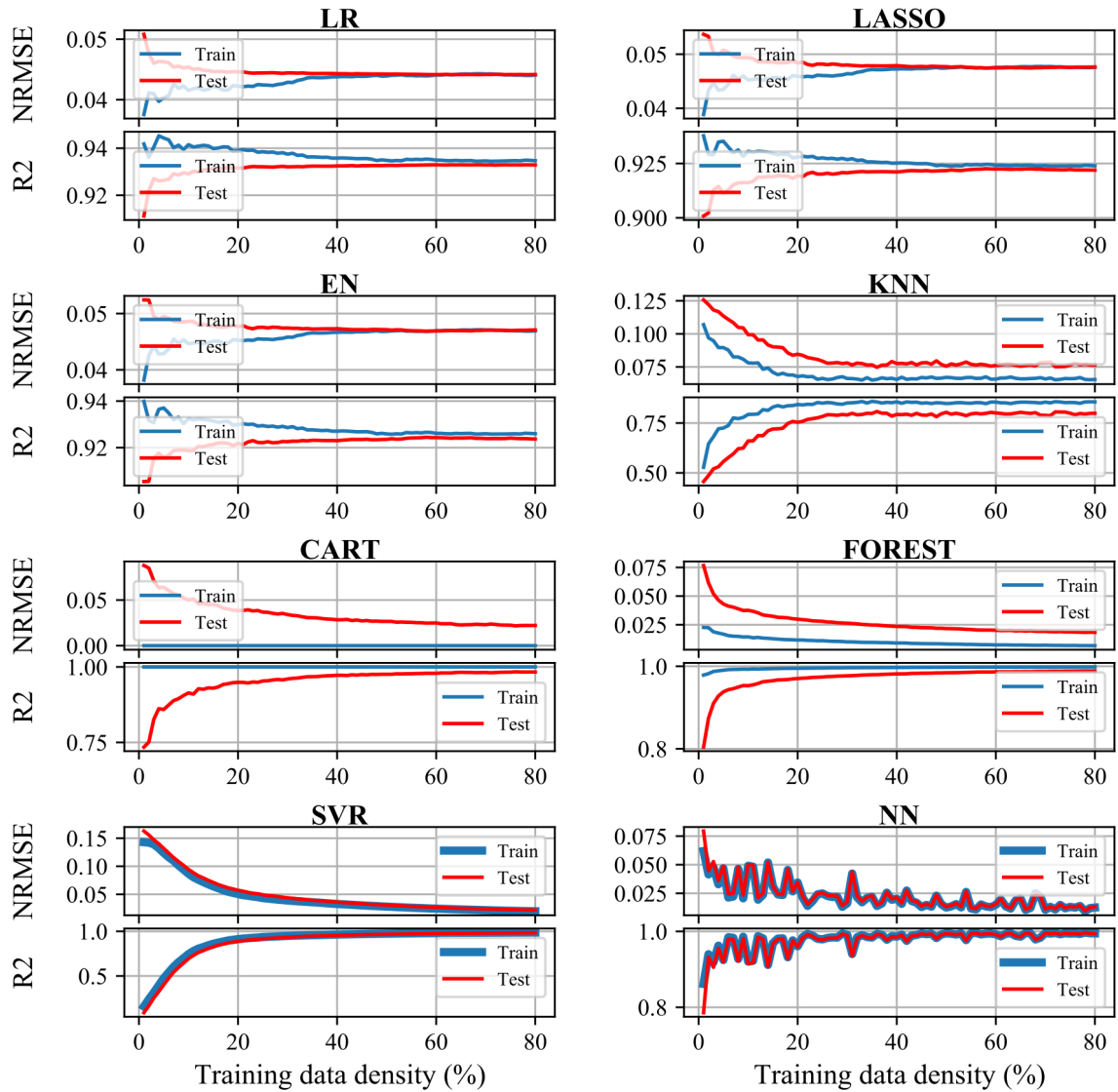

**Figure 5**: Training data density vs. accuracies for the crack propagation of the 2-D composite.

As illustrated in **Figure 6**, $R^2$ for both training and test data approximately converges to 1 as the number of epochs rises. This pattern is also observed for $NRMSE$, which reaches nearly 0. The accuracies show little improvement after 10 epochs; thus, training with epochs higher than 10 proves unnecessary and may lead to overfitting. On the other hand, the batch size, equivalent to the number of subgroups used for computing the cost in a single epoch, leads to higher accuracies as it decreases. Although computing the cost with all data results in higher accuracy, it significantly reduces convergence speed.

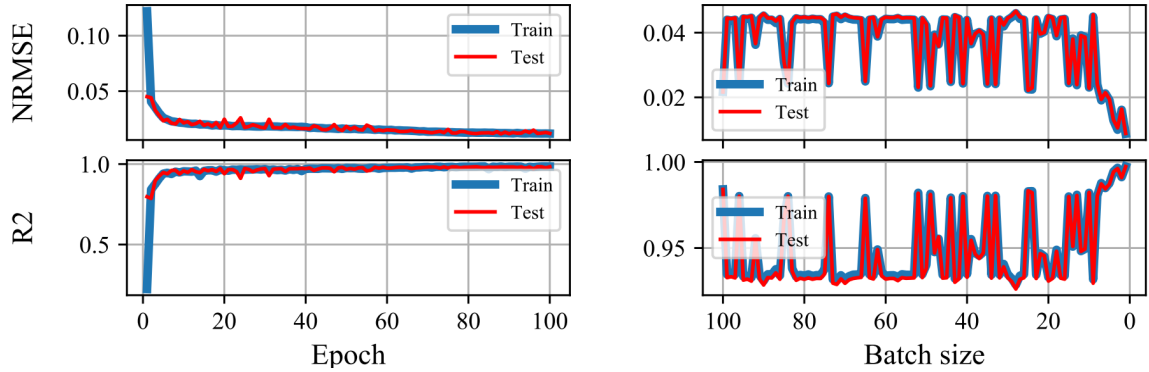

**Figure 6:** Epoch and batch size vs. accuracies of NN for the crack propagation of the 2-D composite

Having normalized the crack propagation values between 0 and 100, **Figure 7** depicts the scattered values for FEM against ML predicted values for each algorithm. The distribution of scattered points around the $y = x$ line indicates the performance of each approach. Hence, it is concluded that KNN and NN are the least and most effective approaches, respectively. Additionally, FOREST, SVR, and CART are the next best-performing algorithms, as confirmed by the boxplot cross-validation in **Figure 8**. For this purpose, a 5-fold criterion has been deployed to evaluate the robustness of the results for each algorithm. The low dispersion of the result in each case vividly indicates the trustworthiness of the result for the ML approach.

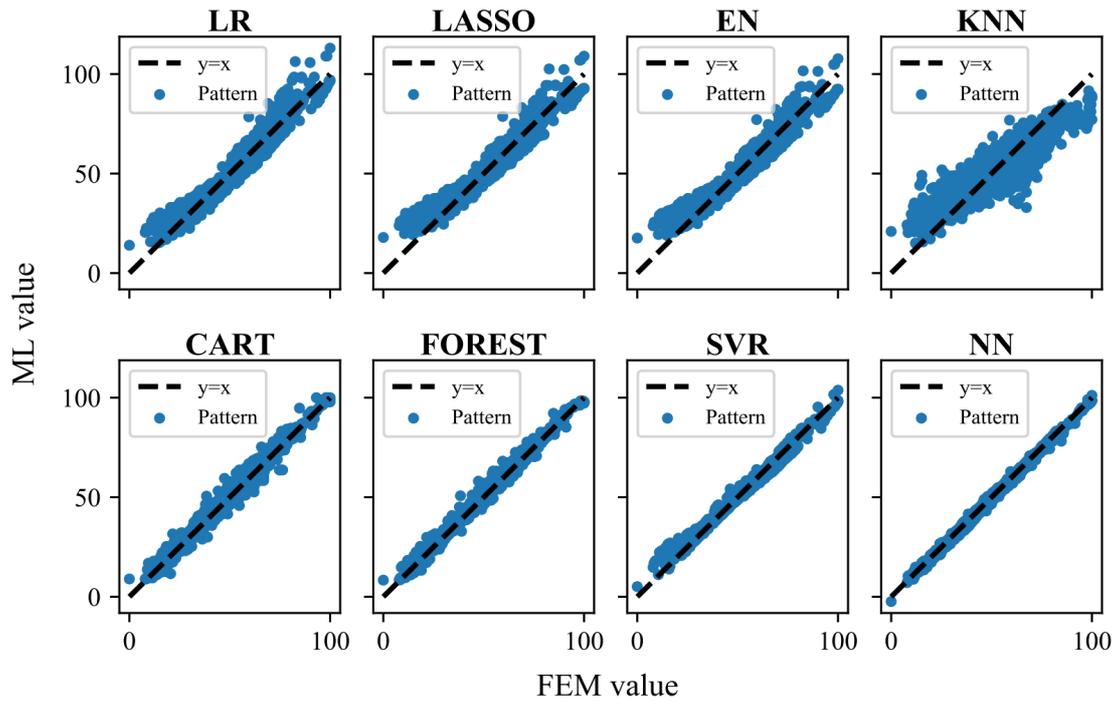

**Figure 7:** FEM values vs. ML values for the crack propagation of the 2-D composite.

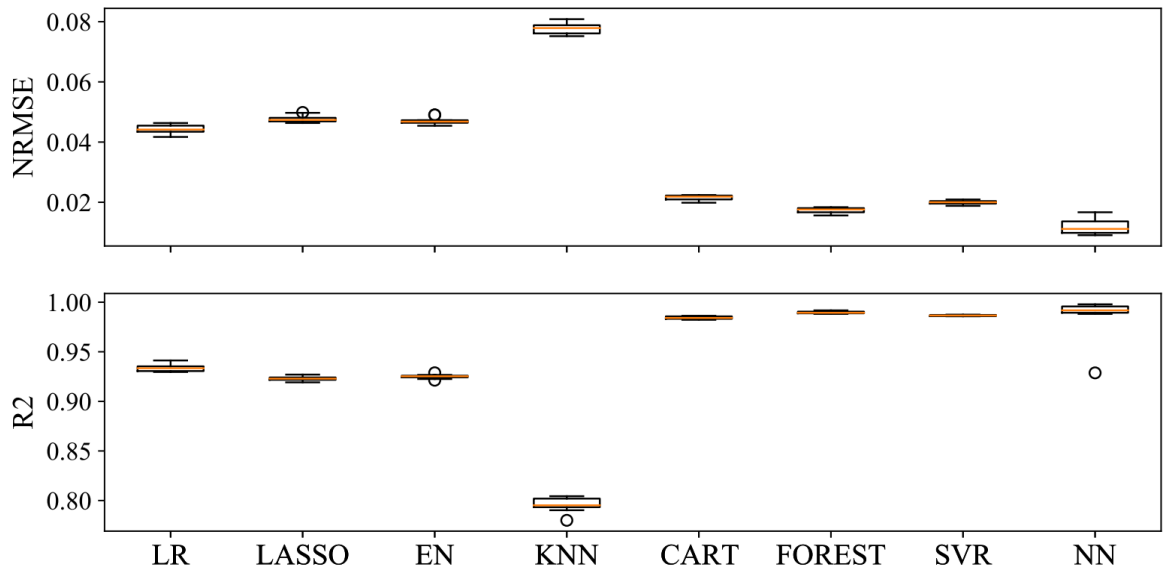

**Figure 8:** 5-fold cross-validation boxplot results for the crack propagation of 2-D composite.

**Table 1** displays the top 10 patterns with the least value of crack propagation based on the best-performing algorithm, which is NN, for the case study. It is evident that the soft elements are positioned around the middle of the beams.

**Table 1:** Top 10 patterns with the least value of crack propagation based on NN for the 2-D composite.

| | | | |
|---|---|---|---|
| $1^{st}$ | | $2^{nd}$ | |
| $3^{rd}$ | | $4^{th}$ | |
| $5^{th}$ | | $6^{th}$ | |
| $7^{th}$ | | $8^{th}$ | |
| $9^{th}$ | | $10^{th}$ | |

## 3.2 Fracture toughness of 2-D model

In this section, the target is the fracture toughness of the 2-D composite. According to **Figure 9,** using the LR algorithm with training consisting of only 10% of generated data, both $R^2$ and NRMSE reach a stable state, equal to 92.5% and 0.055, respectively. Two other linear algorithms, LASSO and EN, produce almost similar results, about 92.5% for $R^2$. The first non-linear method, KNN, still exhibits the weakest performance among all the methods used, with an accuracy of around 80%. CART and FOREST methods maintain good performance, achieving an accuracy of 98% based on 10% of the data. The SVR method undergoes training without fluctuation or overfitting. Although NN demonstrates suitable performance, it undergoes several fluctuations.

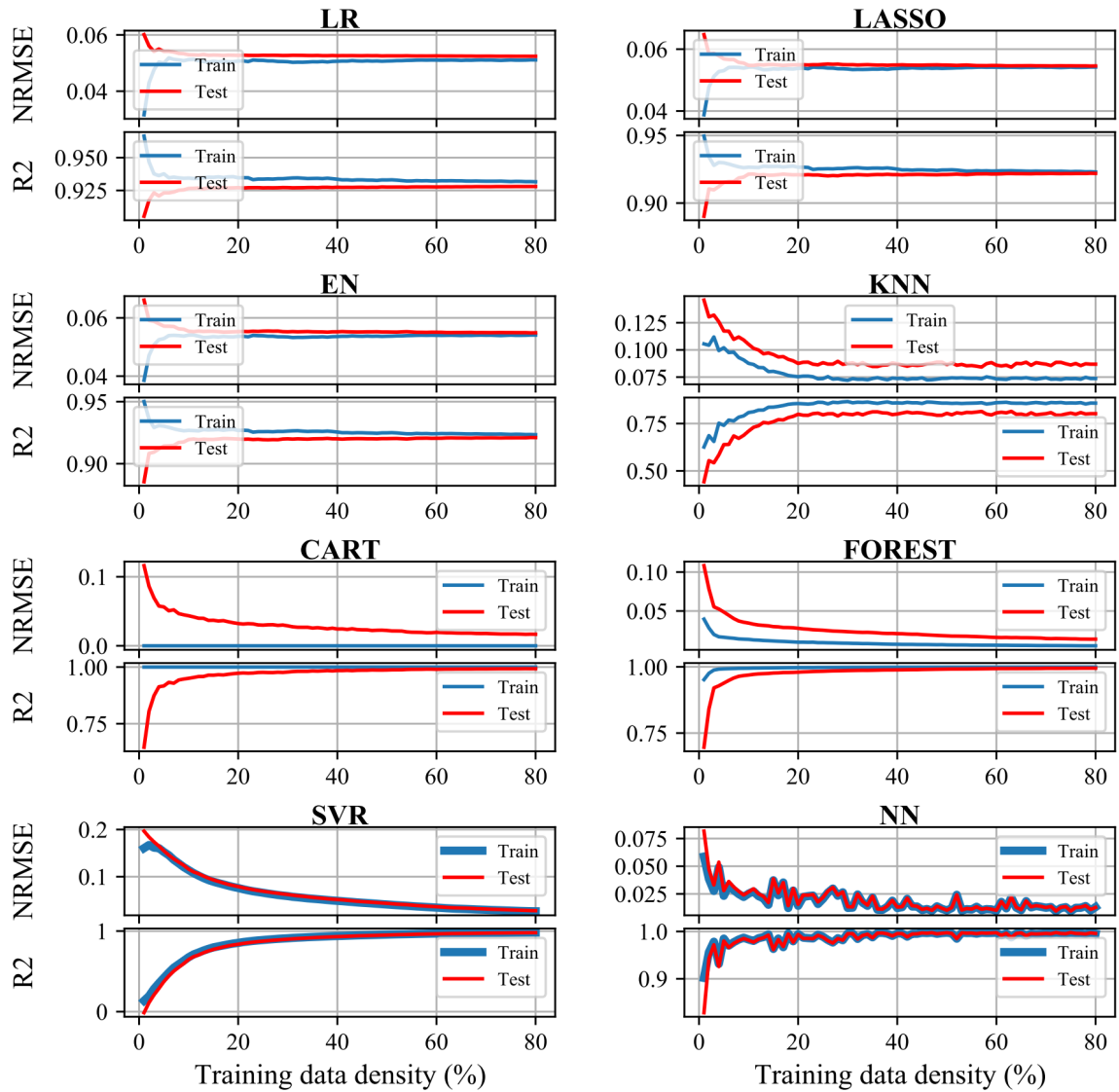

**Figure 9:** Training data density vs. accuracies for the fracture toughness of the 2-D composite.

**Figure 10** illustrates the impact of epoch and batch size parameters on $R^2$ and *NRMSE* for the NN method. After being trained for 20 epochs, the accuracy exceeds 99%, with no noticeable fluctuation. However, changes in batch size, as observed previously, result in numerous fluctuations. According to the figure, the presence of two significant fluctuations suggests that the impact of changing the batch size should always be checked when using the NN method, as it may lead to substantial fluctuations.

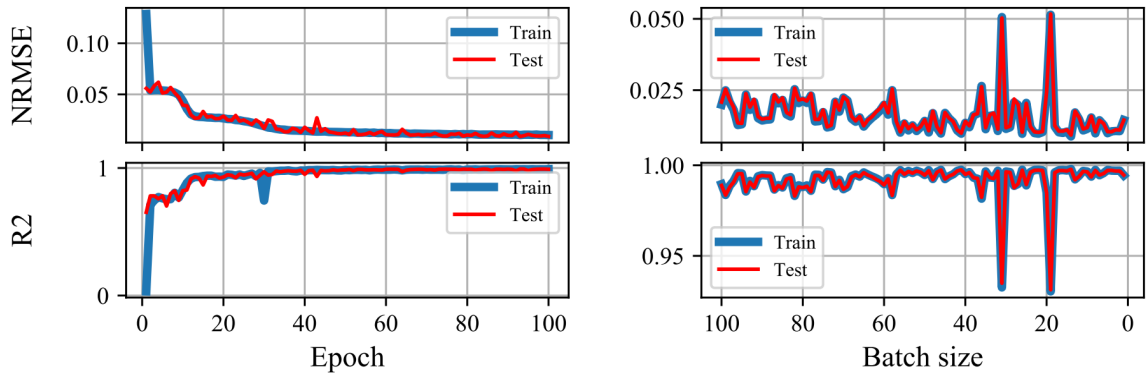

**Figure 10:** Epoch and batch size vs. accuracies of NN for the fracture toughness of the 2-D composite.

**Figure 11** depicts the predictive ability of each method for the fracture toughness value of all 2000 structures available as test data. Similar to the crack propagation of the 2-D composite, NN exhibits the best performance, followed by CART, FOREST, and SVR, respectively. The boxplots demonstarting the error and accuracy of all algorithms based on 5-fold cross-validation are shown in **Figure 12**. The compactness of the boxplots for all algorithms indicates the stability of the results.

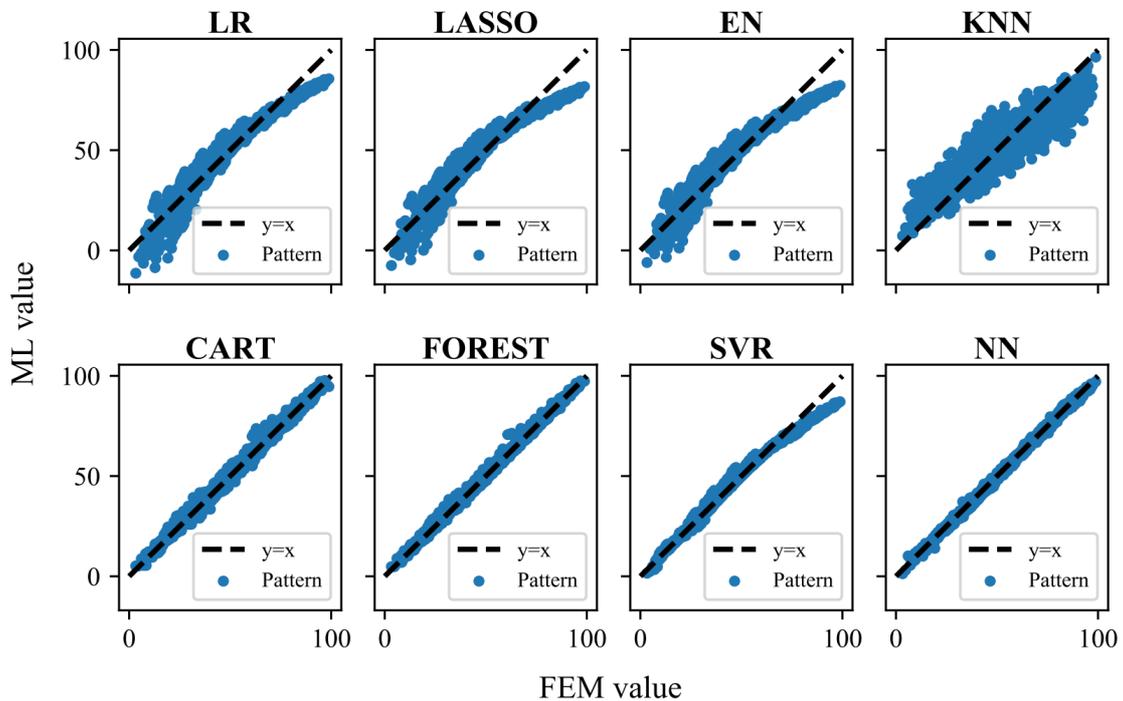

**Figure 11:** FEM values vs. ML values for the fracture toughness of the 2-D composite.

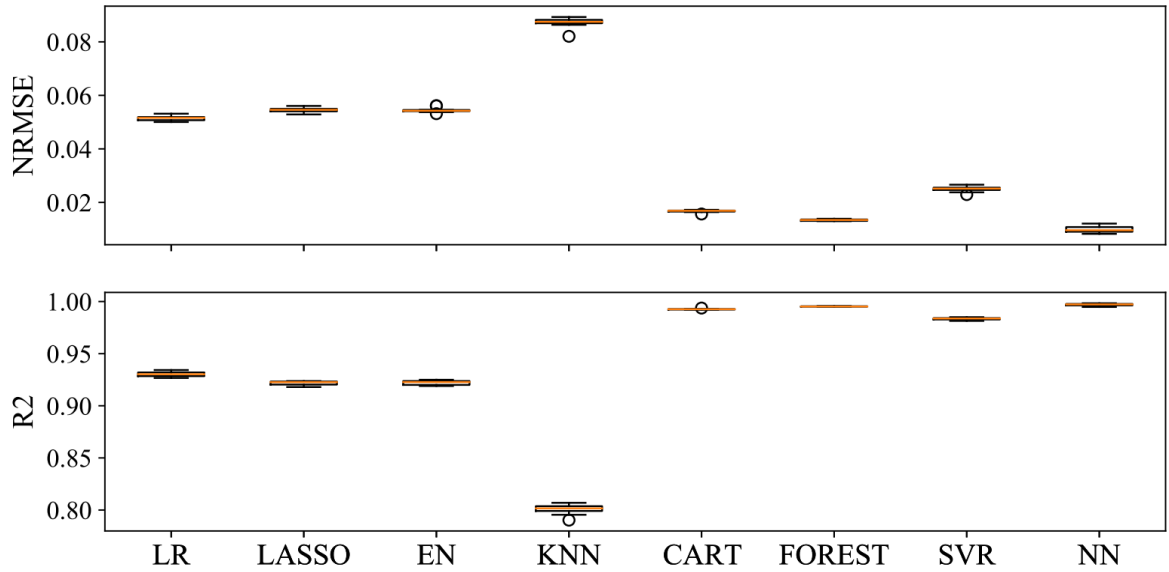

**Figure 12:** 5-fold cross-validation boxplot results for the fracture toughness of the 2-D composite.

Using the NN algorithm as the best-performing, **Table 2** presents the top 10 patterns with the highest fracture toughness, respectively. Examining the patterns, it can be acknowledged that placing soft elements around the crack tip contributes to achieving higher fracture toughness.

**Table 2:** Top 10 patterns with the highest fracture toughness value based on neural networks for the 2-D composite.

| | | | |
|---|---|---|---|
| $1^{st}$ | | $2^{nd}$ | |
| $3^{rd}$ | | $4^{th}$ | |
| $5^{th}$ | | $6^{th}$ | |
| $7^{th}$ | | $8^{th}$ | |
| $9^{th}$ | | $10^{th}$ | |

### 3.3 Crack propagation of 3-D model

This section investigates the target output of the surface of crack propagation in the 3-D composite. The impact of training data density on accuracy and error criteria for all machine learning methods in predicting the crack propagation of the 3-D model is illustrated in **Figure 13**. For LR, despite increasing the training data to 80%, no significant improvement is observed for either criterion. Two other linear methods, LASSO and EN, achieve about 50% and 35% accuracy for $R^2$, respectively. Unlike the other two linear methods, LR shows overlapping accuracy values for the test and training data. The first non-linear method, KNN, reaches a peak of 75% accuracy on the test data, and it is worth noting that the accuracy values on the test and train data do not converge. Meanwhile, CART and FOREST methods demonstrate excellent performance with accuracy convergence close to 99%, based on 20% of all input patterns. The accuracy improvement rate of FOREST is slightly higher than that of CART. The SVR method completes the training cycle without overfitting but with a final accuracy of approximately 75%. NN also exhibits acceptable performance with fluctuations, achieving an accuracy of almost 95% at the end of the process.

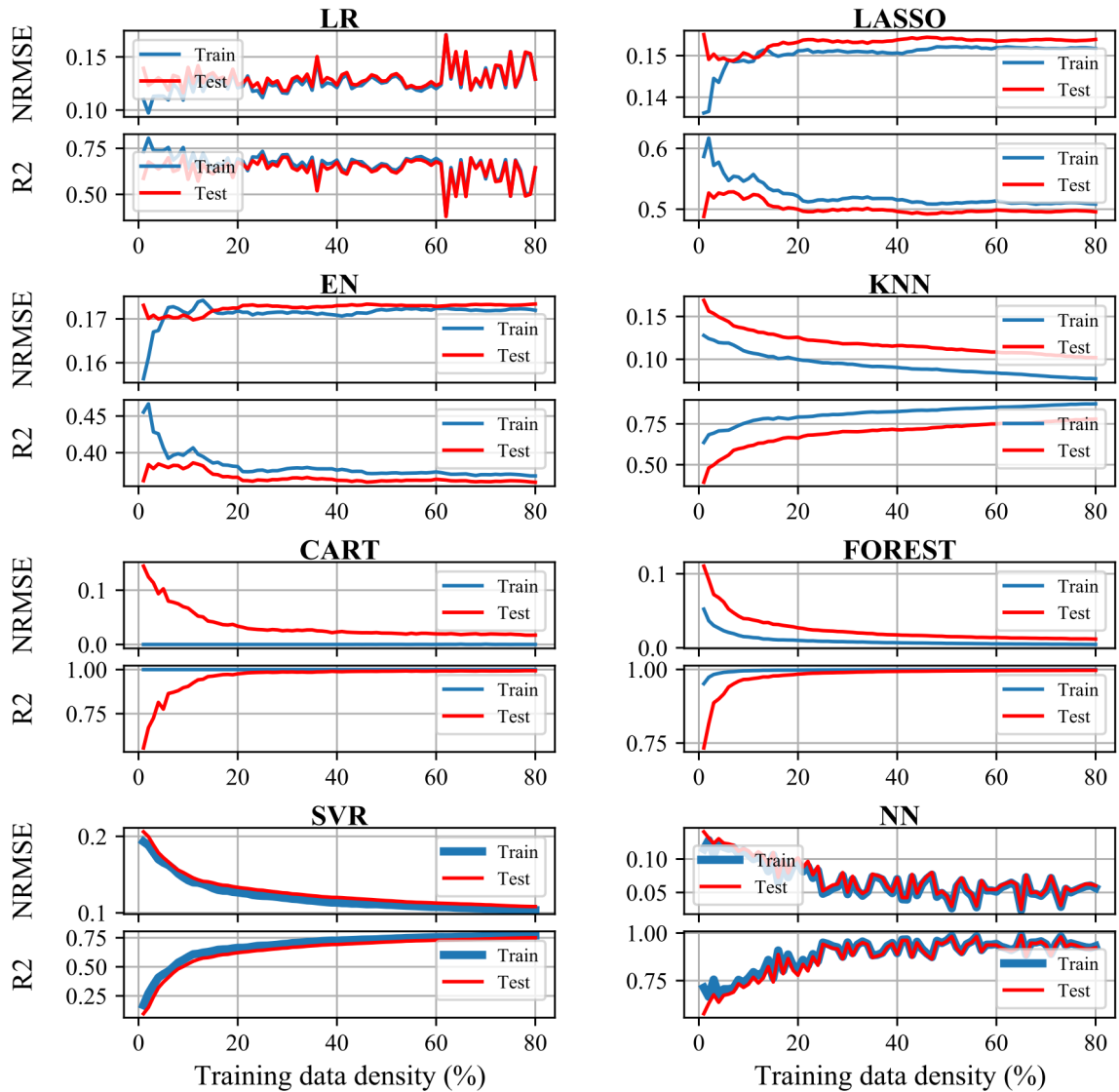

**Figure 13:** Training data density vs. accuracies for the crack propagation of 3-D composite.

**Figure 14** indicates that, after being trained for 40 epochs, the network achieves an accuracy of 95% and an error under 0.05. Therefore, training with a larger number of epochs has little effect on improving accuracy and reducing error. In this problem, using all training data (80% of data) to update weights (Batch size = 1) results in the lowest error (0.02) and highest accuracy (95%).

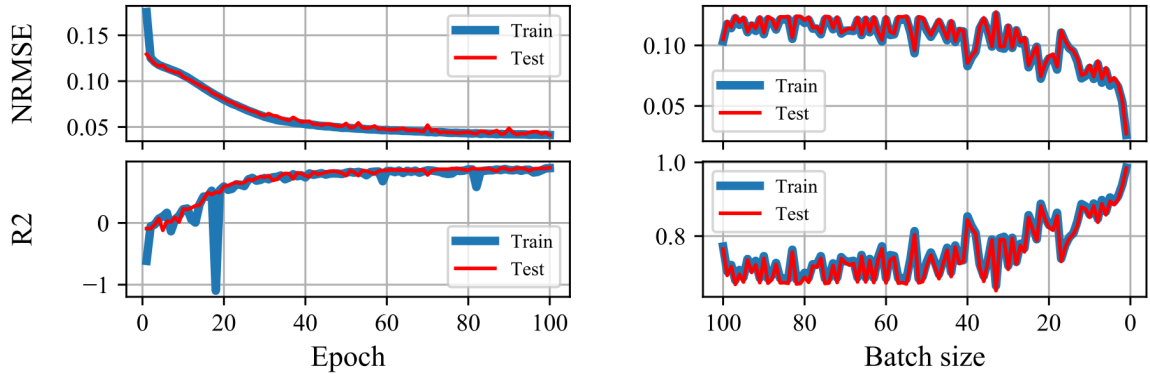

**Figure 14:** Epoch and batch size vs. accuracies of NN for the crack propagation of 3-D composite.

According to **Figure 15**, the FOREST is the best-performing, followed by CART and NN. As depicted in **Figure 16**, the 5-fold cross-validation boxplot for all algorithms, except NN and LR, exhibits relatively good compression. It's worth noting that while NN's compression result may not be the same as CART and FOREST, its average performance is higher compared to the other methods.

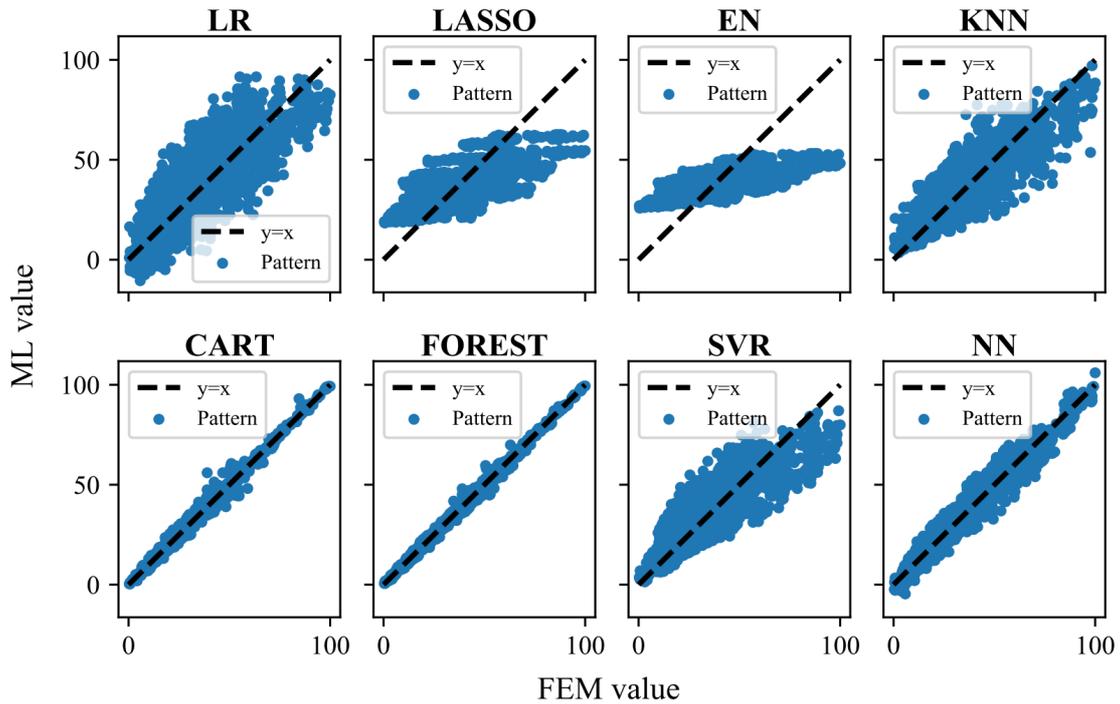

**Figure 15:** FEM values vs. ML values for the crack propagation of 3-D composite.

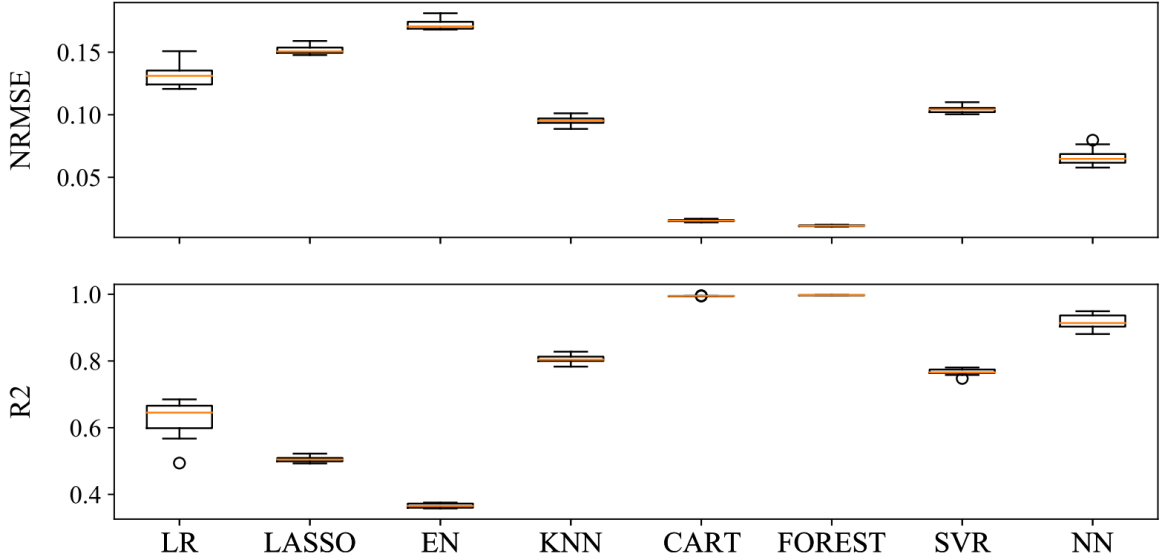

**Figure 16:** 5-Fold cross-validation boxplot results for the crack propagation of 3-D composite.

Using FOREST, the best-performing algorithm, **Table 3** depicts the top 10 layup arrangements with the lowest crack propagation among the 30,240 possible arrangements. Layers, based on fiber orientation, are arranged symmetrically, considering the adhesive surface as the reference line.

**Table 3:** Top 10 layup configurations with the least value of the crack propagation based on FOREST for the 3-D composite.

| Rank | Layup configuration |
|---|---|
| 1st | [90/-45/45/60/-60]s |
| 2nd | [-60/-45/45/60/90]s |
| 3rd | [90/45/30/60/-60]s |
| 4th | [60/45/30/90/-60]s |
| 5th | [60/45/30/-60/90]s |
| 6th | [-60/45/30/60/90]s |
| 7th | [90/60/30/-45/-60]s |
| 8th | [90/60/45/-45/-60]s |
| 9th | [90/60/30/45/-60]s |
| 10th | [60/-45/45/90/-60]s |

## 3.4 Fracture toughness of 3-D model

The target output in this section is the fracture toughness of the 3-D composite, and the effect of training data density on accuracy and error is shown in **Figure 17**. For LR, increasing the training data up to 80% shows no sustainable improvement for $R^2$ and

NRMSE, with eventual values of 25% and 0.175, respectively. LASSO and EN also yield poor results with 22% and 19% for $R^2$, respectively. KNN shows a noticeable discrepancy between the test and train data accuracies, with $R^2$ on the test data peaking at 55%. CART and FOREST methods demonstrate acceptable performance compared to others. Trained with 80% of the data, the CART algorithm achieves an accuracy of about 77% on the test data. Additionally, the accuracy of FOREST on the test and training data is 87% and 98%, respectively. However, SVR achieves an accuracy of 50% when trained with 80% of the data. At the end, the NN approach demonstrates acceptable performance with 82% and 88% accuracy on the test and training data, respectively.

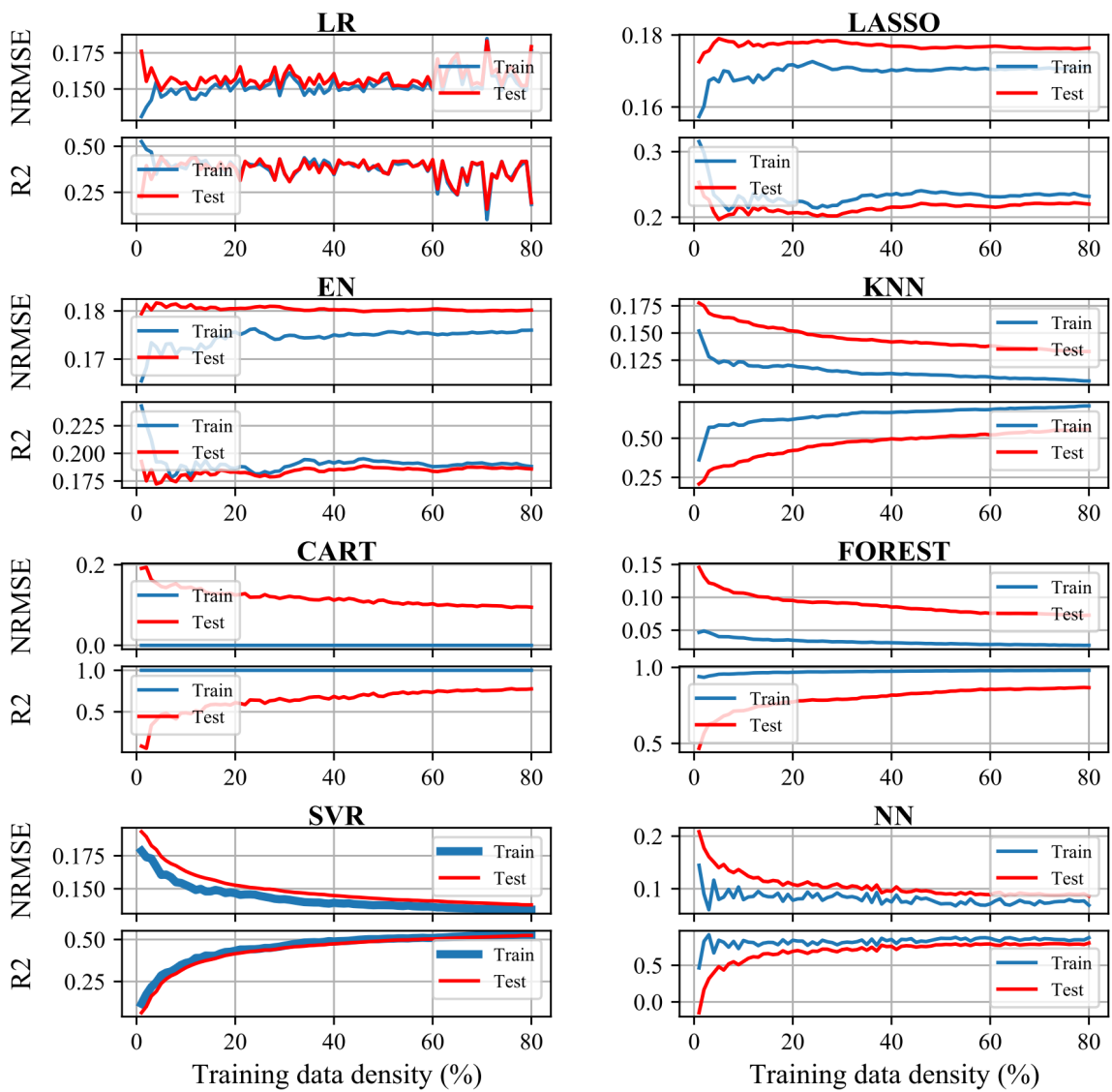

**Figure 17:** Training data density vs. accuracies for the fracture toughness of the 3-D composite.

**Figure 18** illustrates the effect of epoch and batch size on $R^2$ and $NRMSE$. According to the figure, the network reaches the same accuracy value on the test and training data after being trained for 250 epochs. Therefore, training with higher numbers of epochs can cause overfitting. The batch size variation also has a minor effect on the accuracy of the network.

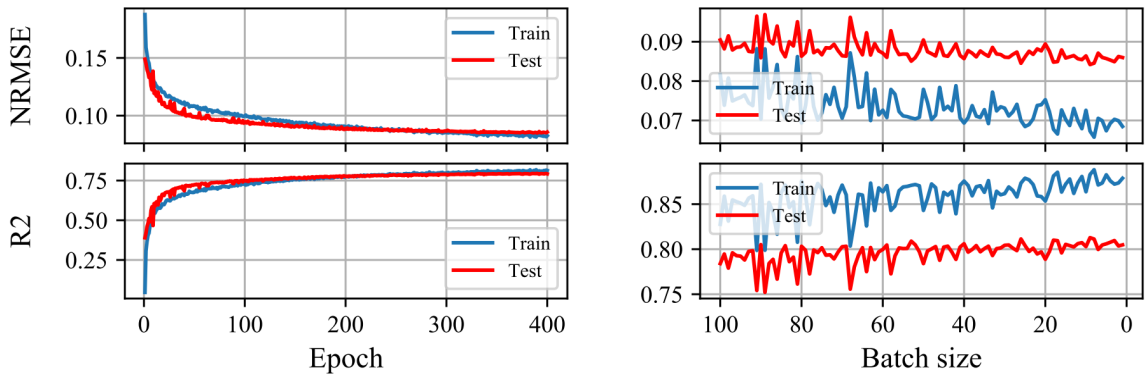

**Figure 18:** Epoch and batch size vs. accuracies of NN for the fracture toughness of the 3-D composite.

Based on **Figure 19**, the FOREST algorithm performs best, and NN and CART are placed in the following ranks, respectively. Additionally, the boxplots shown in **Figure 20** suggest that NN, FOREST, and CART have relatively good compression, indicating the stability of the results.

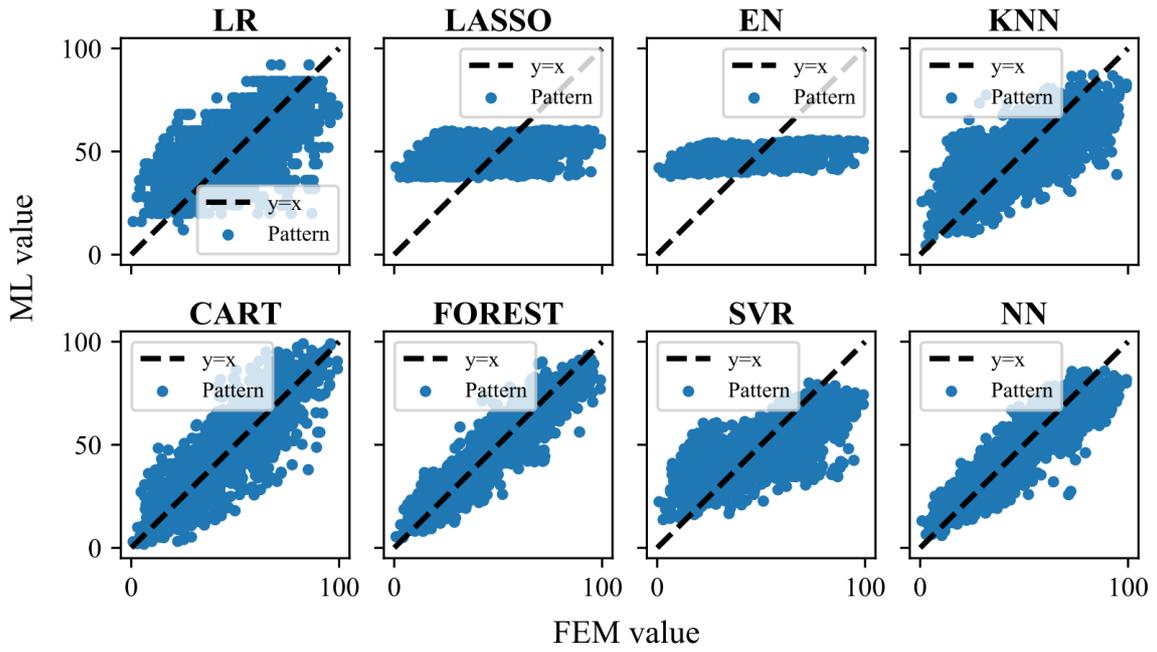

**Figure 19:** FEM values vs. ML values for the fracture toughness of the 3-D composite.

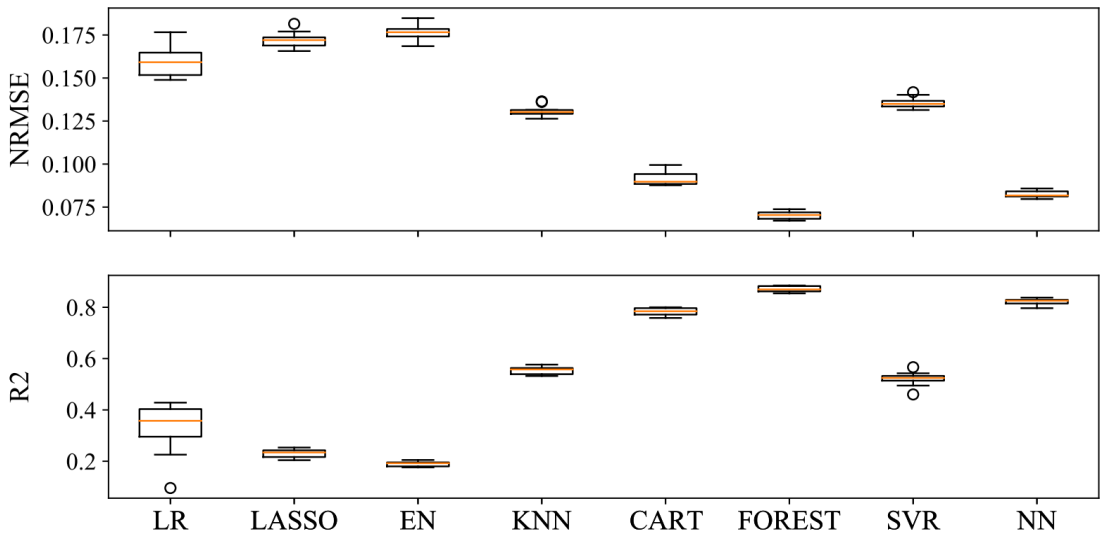

**Figure 20:** 5-Fold cross-validation boxplot results for the fracture toughness of the 3-D composite.

Based on the FOREST (the best performing among all algorithms), **Table 4** depicts the top 10 layup arrangements with the highest fracture toughness values among 30,240 possible arrangements. The layers (based on fiber orientation) are arranged symmetrically, considering the adhesive surface as the reference line.

**Table 4:** Top 10 layup configurations with the highest fracture toughness value based on FOREST for the 3-D composite.

| Rank | Layup configuration |
|---|---|
| 1st | [45/90/-15/60/-45]s |
| 2nd | [45/90/15/60/-45]s |
| 3rd | [90/-30/60/-60/45]s |
| 4th | [45/-60/-15/60/-45]s |
| 5th | [60/90/0/45/-45]s |
| 6th | [90/-30/60/-60/-45]s |
| 7th | [45/90/15/-60/-45]s |
| 8th | [45/-60/15/60/-45]s |
| 9th | [-60/60/15/45/-30]s |
| 10th | [90/45/15/-60/30]s |

## 4. Discussion

While the deployed ML algorithms show promising performance in predicting FEM values for the behaviors of 2-D and 3-D composites, the true potential of using ML to design and optimize composites lies in modeling much larger systems. ML can serve as a substitute for physics-based approaches, especially when dealing with vast design spaces and unmanageable amounts of data. This research demonstrates that mechanical properties can be predicted using various ML algorithms even with a minimal amount of training data. Evaluating more extensive systems for 2-D composites (e.g., 4 x 128 elements) or 3-D composites with more laminas is feasible, achieving optimal designs with limited training datasets. Recent research has explored optimization approaches for composite structures, which can now be physically constructed using modern 3-D printing technology.

It's important to note that this work aims to showcase the application of ML to a composite design problem while identifying the best-performing ML algorithm for the case study. The computational cost for finite element simulation of a single 2-D model is approximately 32 seconds, requiring about 1.15 million seconds to calculate the mechanical properties of 35,960 patterns. Similarly, a single 3D model structure takes about 199 seconds, and it takes about 6.02 million seconds to examine 30,240 layups for the 3-D model. All deployed algorithms took less than 10 seconds to be trained by 80% of the data, except for NN, which took 460 seconds and 895 seconds for 2-D and 3-D model cases, respectively. NN and FOREST emerged as the best-performing ML algorithms for 2-D and 3-D models, working significantly faster than FEM—about 2,500 times faster for NN and 760,000 times faster for FOREST. However, the computational cost of generating training data by FEM should not be overlooked. In summary, for optimal design exploration in a vast design space, it is recommended to conduct proper sampling, use ML to discern general trends, and validate results through experimental or computational methods.

## 5. Conclusion

In this research, ML was employed to predict the mechanical behavior of 2-D and 3-D composites as beams in the DCB test. High-performance patterns and arrangements were identified to prevent crack growth and increase fracture toughness. The study evaluated the effects of different training parameters, such as data density, epochs, and batch size, for various algorithms. Cross-validation demonstrated the independence of results from data selection for training and testing. The best-performing algorithm, with superior capabilities, positions ML as a promising tool for studying larger systems in designing and optimizing composite patterns and layups.

Additionally, integrating 3D printing technology can significantly enhance the practical application of this study's findings. 3D printing allows for the precise fabrication of composite structures based on ML-optimized designs. This capability not only enables the efficient prototyping and validation of these designs but also ensures the achievement of optimal configuration designs that are tailored to specific performance requirements. By leveraging 3D printing, researchers can efficiently implement and test the most effective composite layouts identified through ML, facilitating advancements in material performance and structural integrity.

## Funding

The authors received no financial support for the research, authorship, and/or publication of this article.

## Declaration of competing interest

The authors declare that they have no known competing financial interests or personal relationships that could have appeared to influence the work reported in this paper.

## Data availability

Data will be made available on request.